\documentclass[aps,showpacs,preprintnumbers,amsmath, amssymb]{revtex4}

\usepackage{float}
\usepackage{graphics,epsfig}
\usepackage{graphicx}
\usepackage{dcolumn}
\usepackage{bm}
\usepackage{amsmath}
\usepackage{mathrsfs}

\begin{document}
\baselineskip=0.8 cm

\title{{\bf A unified hoop conjecture for black holes and horizonless compact stars}}
\author{Yan Peng$^{1,2}$\footnote{yanpengphy@163.com}}
\affiliation{\\$^{1}$ School of Mathematical Sciences, Qufu Normal University, Qufu, Shandong 273165, China}
\affiliation{\\$^{2}$ Center for Gravitation and Cosmology, College of Physical Science
and Technology, Yangzhou University, Yangzhou 225009, China}

\vspace*{0.2cm}
\begin{abstract}
\baselineskip=0.6 cm
\begin{center}
{\bf Abstract}
\end{center}

We propose a unified version of hoop conjecture
valid for various black holes and horizonless compact stars.
This conjecture is expressed by the mass to circumference ratio
$4\pi M_{in}/C\leqslant 1$, where C is the circumference of
the smallest ring that can engulf the object in all azimuthal directions
and $M_{in}$ is the mass within the engulfing sphere.

\end{abstract}

\maketitle
\newpage
\vspace*{0.2cm}

\section{Introduction}

The famous hoop conjecture introduced almost five decades ago
asserts that the existence of black hole horizons is characterized
by the mass and circumference relation $4\pi \mathcal{M}/C\geqslant 1$ \cite{hc1,hc2}.
Here C is the circumference of the smallest ring that can engulf the
black hole in all azimuthal directions and $\mathcal{M}$ is usually interpreted
as the asymptotically measured total ADM mass of black holes \cite{hc3}-\cite{ahc9}.

For horizonless compact stars, the curved spacetimes
should be characterized by the relation $4\pi \mathcal{M}/C< 1$ \cite{hc1,hc2}.
However, if $\mathcal{M}$ is still
interpreted as the total ADM mass,
this relation $4\pi \mathcal{M}/C< 1$
is violated in the background of horizonless charged compact objects \cite{hc19,hc20}.
In fact, Thorne has not provided an explicit definition for the mass
term $\mathcal{M}$ in the mass to circumference ratio \cite{hc1}.
And it has been proved that the relation $4\pi \mathcal{M}/C< 1$
holds in horizonless spacetimes when $\mathcal{M}$ is interpreted
as the mass contained within the engulfing sphere (not as
the total ADM mass) \cite{hc21,hc22}.

It is natural to assume that a unified hoop conjecture may be
$4\pi \mathcal{M}/C\geqslant 1$ for black holes and $4\pi \mathcal{M}/C< 1$ for
horizonless compact stars, where
$\mathcal{M}$ is the total ADM mass or the mass contained in the
engulfing sphere \cite{hc23}. In the background of horizonless stars,
the calculations in \cite{hc21} imply that the term $\mathcal{M}$ should be the mass
contained in the sphere and cannot be the total ADM mass.
However, in the black hole spacetimes, Hod found that
the hoop conjecture holds if $\mathcal{M}$ is the
ADM mass and cannot be the mass contained in the sphere \cite{hc23}.
So it seems that there is no unified hoop conjecture
valid for both black holes and compact stars.

In this paper, we propose another version of unified hoop conjecture,
which holds for Schwarzschild black holes, neutral Kerr black holes,
RN black holes, Kerr-Newman black holes and horizonless charged compact stars.
It may be a general property for both black holes and horizonless compact stars.

\section{Test the unified hoop conjecture}

In the background of horizonless charged compact stars,
it has recently been proved that the relation $4\pi \mathcal{M}/C< 1$
is violated when $\mathcal{M}$ is interpreted as the total ADM mass
and the relation holds if $\mathcal{M}=M_{in}$,
where $M_{in}$ is the mass contained within the engulfing sphere \cite{hc19,hc20,hc21,hc22}.
In addition, the numerical data
in \cite{hc23} implies that $4\pi M_{in}/C\leqslant 1$ may be a general property in black hole
spacetimes. With these facts, we propose a unified version of hoop conjecture
\begin{eqnarray}\label{AdSBH}
4\pi M_{in}/C\leqslant 1,
\end{eqnarray}
where C is the circumference of
the smallest ring that can engulf the object in all azimuthal directions
and $M_{in}$ is the mass within the engulfing sphere.
The relation (1) is natural according to the idea that
the ring of the engulfing sphere is related to the mass within the sphere
(not the mass in the total spacetime).

In the following, we give examples supporting the bound (1)
following studies in \cite{hc19,hc20,hc21,hc22,hc23}. A Kerr-Newman black hole spacetime is
described by the curved line element \cite{hc23}
\begin{eqnarray}\label{AdSBH}
ds^{2}&=-\frac{\Delta-a^2sin^2\theta}{\rho^2}dt^2+\frac{\rho^2}{\Delta^2}dr^2
-\frac{2a sin^2\theta(2Mr-Q^2)}{\rho^2}dtd\phi+\rho^2d\theta^{2}
+\frac{(r^2+a^2)^2-a^2 \Delta sin^2\theta}{\rho^2}sin^2\theta d\phi^{2},
\end{eqnarray}
where M is the ADM mass, Q is the electric charge, $J=Ma$
is the angular momentum, $\Delta=r^2-2Mr+Q^2+a^2$ and $\rho^2=r^2+a^2cos^2\theta$.
Black hole horizons are expressed as
\begin{eqnarray}\label{AdSBH}
r_{\pm}=M\pm(M^2-Q^2+a^2)^{1/2},
\end{eqnarray}
which are determined by the zeros of the metric function $\Delta(r)=0$.

Setting $dt=dr=d\theta=0,\theta=\frac{\pi}{2}$,
we get the relation
\begin{eqnarray}\label{AdSBH}
ds=\frac{r_{+}^2+a^2}{r_{+}}d\phi.
\end{eqnarray}

Also taking $\Delta\phi=2\pi$,
we obtain the equatorial circumference
\begin{eqnarray}\label{AdSBH}
C=2\pi\frac{r_{+}^2+a^2}{r_{+}}.
\end{eqnarray}

For Schwarzschild black holes, there is $Q=0$ and $a=0$.
The mass to circumference ratio is
\begin{eqnarray}\label{AdSBH}
\frac{4\pi M_{in}}{C}=\frac{4\pi M}{C}=\frac{4\pi M}{2\pi r_{+}}=\frac{4\pi M}{2\pi (2M)}=1.
\end{eqnarray}

For RN black holes with $Q\neq0$ and $a=0$,
the ratio satisfies
\begin{eqnarray}\label{AdSBH}
\frac{4\pi M_{in}}{C}=\frac{4\pi (M-\frac{Q^2}{2r_{+}})}{2\pi r_{+}}
=\frac{4Mr_{+}-2Q^2}{2r_{+}^{2}}=\frac{4M(M+\sqrt{M^2-Q^2})-2Q^2}{2(M+\sqrt{M^2-Q^2})^{2}}=1.
\end{eqnarray}

The neutral Kerr black holes correspond to $Q=0$ and $a\neq0$.
It yields the relation
\begin{eqnarray}\label{AdSBH}
\frac{4\pi M_{in}}{C}=\frac{4\pi M}{2\pi \frac{r_{+}^2+a^2}{r_{+}}}=\frac{2Mr_{+}}{r_{+}^2+a^2}=\frac{2M(M+\sqrt{M^2-a^2})}{(M+\sqrt{M^2-a^2})^2+a^2}=1.
\end{eqnarray}

In the background of Kerr-Neuman black holes, both charge and
rotating parameters are nonzero as $Q\neq0$ and $a\neq0$.
The equatorial circumference is $C=2\pi \frac{r_{+}^2+a^2}{r_{+}}$
and the mass contained within the black hole horizon
is $M_{in}=M-\frac{Q^2}{4r_{+}}[1+\frac{r_{+}^2+a^2}{ar_{+}}arctg(\frac{a}{r_{+}})]$ \cite{mass}.
In this case of Kerr-Neuman black holes, numerical data in \cite{hc23} implies an upper bound
on the ratio
\begin{eqnarray}\label{AdSBH}
4\pi M_{in}/C\leqslant 1.
\end{eqnarray}

The horizonless charged compact star satisfies the inequality \cite{hc21,hc22}
\begin{eqnarray}\label{AdSBH}
4\pi M_{in}/C< 1.
\end{eqnarray}

According to relations (6-10), we propose
the unified hoop conjecture (1), which may be a general property for
both black holes and horizonless compact stars.

\section{Conclusions}

We proposed a unified hoop conjecture,
which is valid for various black holes and horizonless compact stars.
Our conjecture is that the mass and circumference relation
$4\pi M_{in}/C\leqslant 1$ holds, where C is the circumference of
the smallest ring that can engulf the object in all azimuthal directions
and $M_{in}$ is the gravitating mass within the engulfing sphere.
Our statement is in accordance with the idea that
the ring of the engulfing sphere should be related to the mass within the sphere
(not the mass in the total spacetime).

\begin{acknowledgments}

This work was supported by the Shandong Provincial Natural Science Foundation of China under Grant
No. ZR2018QA008. This work was also supported by a grant from Qufu Normal University
of China under Grant No. xkjjc201906.

\end{acknowledgments}

\end{document}